\documentclass{article}
\usepackage{graphicx}
\usepackage{epsfig}
\title{Low frequency current noise of the single-electron
  shuttle}
\author{A. Isacsson$^{1,2}\footnote{Email:\texttt{isac@fy.chalmers.se}}$ and T. Nord$^{2}$}

\date{}
\begin{document}

\maketitle
\begin{center}
{\it $^1$Department of Physics, Yale University, \\
P.O. Box 208120, New Haven CT 06520-8120\\
  $^2$ Department of Applied
  Physics, Chalmers University of
  Technology \\ and G\"oteborg University, SE-412 96 G\"oteborg,
  Sweden}
  \end{center}
\vspace*{1cm}

\begin{abstract}
  Coupling between electronic and mechanical degrees of freedom in a
  single electron shuttle system can cause a mechanical instability
  leading to shuttle transport of electrons between external leads. We
  predict that the resulting low frequency current noise can be
  enhanced due to amplitude fluctuations of the shuttle oscillations.
  Moreover, at the onset of mechanical instability a pronounced peak
  in the low frequency noise is expected.
\end{abstract}

\newpage

\section{Introduction}
Charge transport in nanostructures is a major research area, both
theoretically and experimentally. Apart from the average current
flowing through a structure in response to an applied external
field, fluctuations and correlations in time of this current are
of interest. By studying the current noise power spectral density
(PSD), information about the charge transport process can be
extracted that may not be accessible through studies of the
average current alone.\cite{Blanter} An example of this is shot
noise, arising due to the discreteness of the charge carriers,
(electrons). \cite{a:93_hanke}

In Nanoelectromechanical Systems (NEMS)~\cite{roukes}, mechanical
degrees of freedom affect and/or are affected by charge transport
through the device. One such system is the single electron shuttle
system~\cite{a:98_gorelik}, also known as the
nanoelectromechanical single electron transistor (NEM-SET). The
system consists of a metallic grain embedded in an elastic
material between two bulk leads, forming a mechanically soft
Coulomb blockade double junction. Since current through the system
is accompanied by charging of the grain, interplay between the
Coulomb forces and the mechanical degrees of freedom can lead to
self-oscillations of the grain, which in turn, supports charge
transport through shuttling of electrons between the leads. Much
work, both
experimental~\cite{a:98_erbe,a:02_scheible,a:00_park,a:02_nagano,a:99_tuominen}
as well as
theoretical~\cite{a:98_gorelik,a:98_isacsson,a:01_nishiguchi,a:02_nord,a:04_nord_isacsson,a:99_weiss,a:02_nishiguchi,
a:01_boese,a:02_fedorets1,a:02_armour,a:02_fedorets2,a:03_novotny,a:03_mccarthy,karsten,a:01_gorelik,a:03_shekhter}
has been reported in this field.

In this paper, the noise spectrum of the single-electron shuttle
is studied in the limit of weak electromechanical coupling. It is
found that the onset of mechanical vibrations is accompanied by a
peak in the low frequency PSD. Hence, by measuring the noise, it
can be determined whether or not the grain is oscillating. This is
an important result since direct detection of any high frequency
mechanical motion is problematic with present day experimental
techniques.

\section{Model System}
\begin{figure}
  \hspace*{.9cm}\includegraphics[width=5cm]{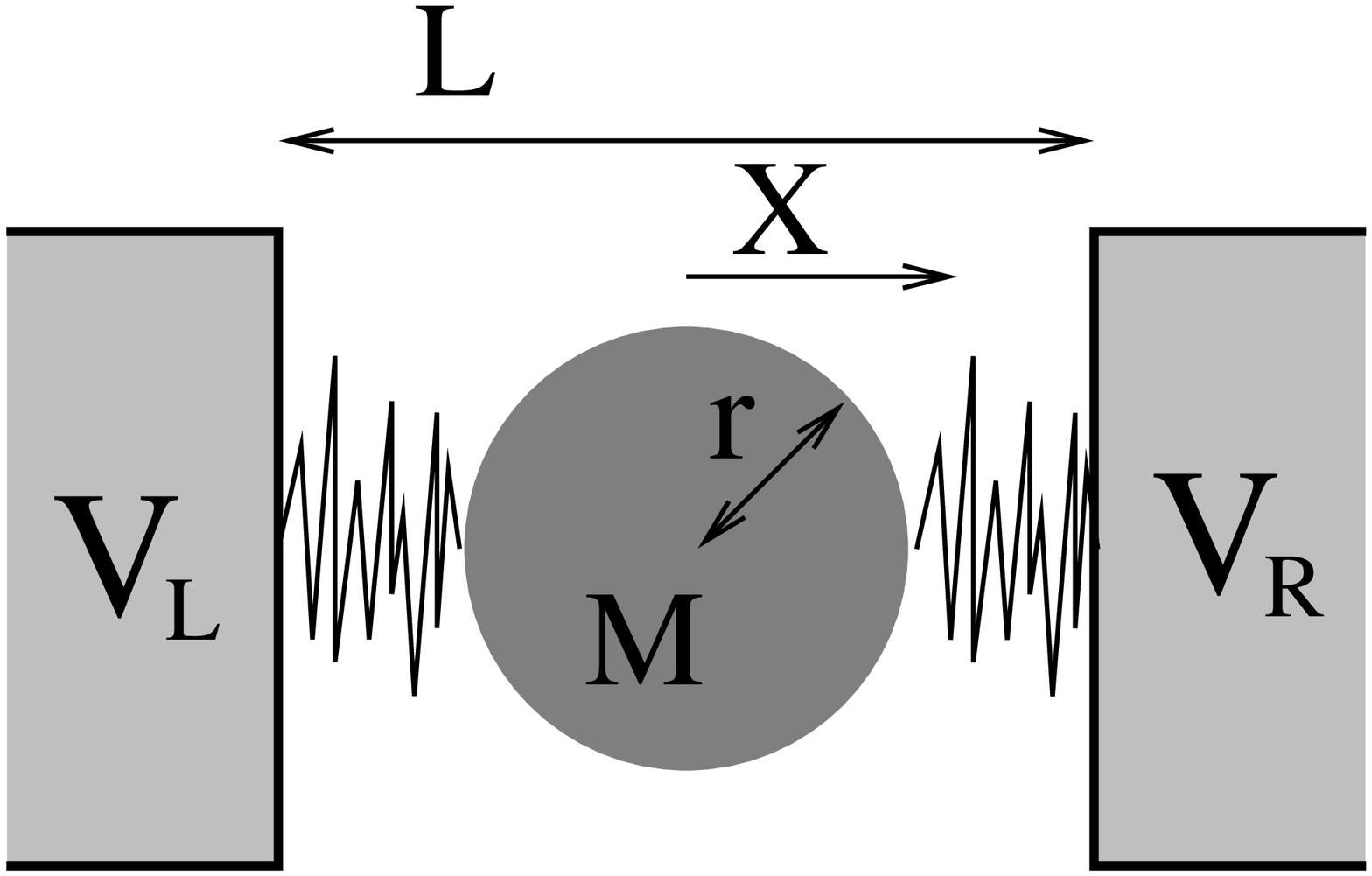}\hspace*{1cm}
  \includegraphics[width=5cm]{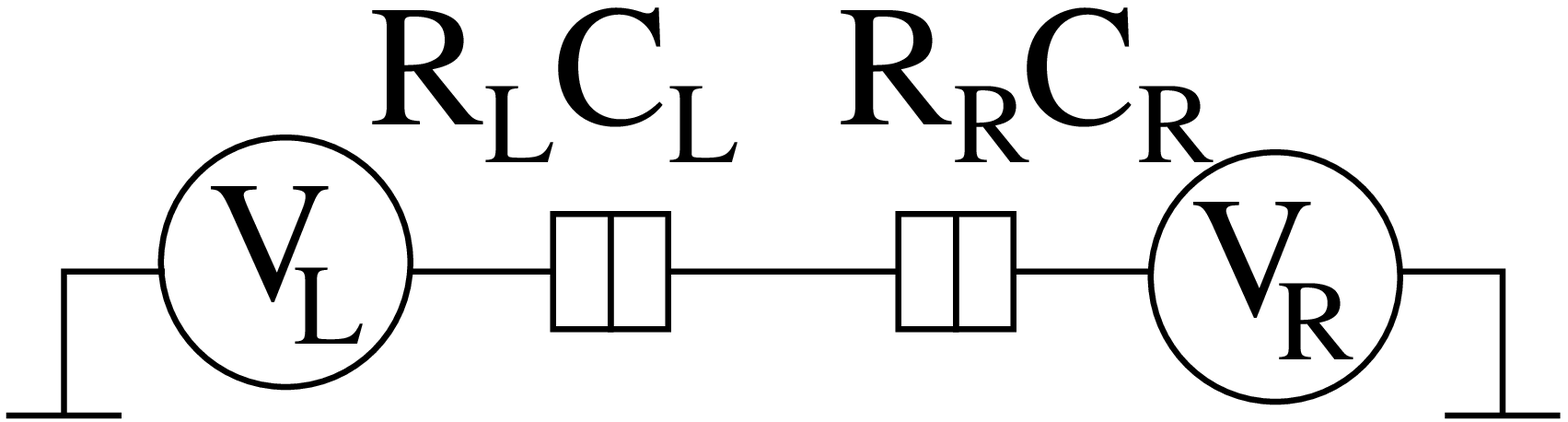}\\
  (a)\hspace*{7.0cm}(b)
  \caption{ \label{fig:model_system}Single electron shuttle. (a)
    A metallic grain of mass $M$ and radius $r$ placed between two
    leads separated by a distance $L$.  The displacement of the
    grain from the center of the system is labelled $X$. The grain
    is connected to the leads via insulating elastic materials. The leads are biased to the potentials $V_L$
    and $V_R$. (b)
    Equivalent circuit of the system. The tunneling resistances and
    capacitances of the left and right junctions are $R_L$, $R_R$,
    $C_L$, and $C_R$. }
\end{figure}

A schematic picture of the system is shown in \mbox{fig.
  \ref{fig:model_system}a}. A metallic grain of mass $M$ and radius
$r$ is placed between two leads, separated by a distance $L$, via
elastic, insulating materials. When a bias voltage $V\equiv
(V_L-V_R)$ is applied between the leads, electron transport occur
by sequential, incoherent, tunneling between the leads and the
grain. The system can also lower its electrostatic energy by
altering the grain position $X$.

The electron transport in the system is described using the notion
of the equivalent circuit of \mbox{fig.
  \ref{fig:model_system}b} characterized by
resistances and capacitances $R_{L,R}$ and $C_{L,R}$ which both
depend on the grain position $X$:
$$
R_{L,R}(X)=R_0^{L,R}\exp(\pm X/\lambda),\,\, C_{L,R}(X) =
{C_0^{L,R}}/({1 \pm {X}/{A_{L,R}}}). $$ Here $\lambda$ is the
characteristic length scale for tunneling and $C_0^{L,R}$ are the
capacitances of the left and right junctions when the grain is at
$X=0$. The coefficients $A_{L,R}$ are typical capacitance length
scales for the left and right tunnel junction respectively. The
grain self-capacitance $C_0$ is also included in the model.

If $E_C \gg \hbar/R(X)C(X)$, where $E_C = e^2/2C$, tunneling is
well described by the ``orthodox'' theory of Coulomb
blockade~\cite{kulik_shekhter} and the tunneling rates through the
left(right) junction is
\begin{equation}
  \Gamma_{L,R}^\pm(X,V,Q) = \frac{\Delta
    G_{L,R}^\pm(X,V,Q)}{e^2R_{L,R}(X)} \frac{1}{1 -
    e^{-\beta\Delta G_{L,R}^\pm(X,V,Q)}
    }.\label{eq:rates}
\end{equation}
Here $\beta$ is the inverse temperature and $\Delta
G_{L,R}^\pm(X,V,Q)$ is the decrease of free energy as the event
$(Q,Q_{L,R}) \to (Q \pm e, Q_{L,R} \mp e)$ occurs. The charges
$Q_{L,R}$ and $Q$ denote the charges accumulated on the left and
right leads and the excess charge on the grain. The free energy
decrease $\Delta G_{L,R}^\pm(X,V,Q)$ can be expressed using the
equivalent circuit model of \mbox{fig.
  \ref{fig:model_system}b}.

Considering the grain motion to be classical and one dimensional
we have Newton's equation $M\ddot{X}=F_{\rm Ext.}(X,V,Q)$ for the
grain displacement. The external force $F_{\rm Ext.}(X,V,Q)$
acting on the grain includes an elastic force $F_k(X)$, a
dissipative force $F_{\gamma}(\dot{X})$, an electric force
$F_{\epsilon}(X,Q,V)$ and a vdW force $F_{vdW}(X)$ yielding the
equation of motion
\begin{equation}
  M\ddot{X} = F_k(X) + F_{\gamma}(\dot{X}) + F_{\epsilon}(X,Q,V) +
  F_{vdW}(X).\label{eq:Newton}
\end{equation}
For the elastic force we use a phenomenological non-linear
potential~\cite{a:04_nord_isacsson} acting as a harmonic well with
spring force constant $k$ for small displacements but that
diverges as a Lennard-Jones potential (12th power) at the
positions $X_L$ and $X_R$:
$$
F_k(X) = \frac{a}{(X+X_L)^{13}} + \frac{b}{(X-X_R)^{13}},
$$
where
$$a = - \frac{k}{13} \frac{(X_0-X_R)(X_0+X_L)^{14}}{X_L+X_R},
  \,\,
  b = \frac{k}{13} \frac{(X_0+X_L)(X_0-X_R)^{14}}{X_L+X_R}.$$
The dissipative force is modelled as a viscous damping term
$F_\gamma(\dot{X}) = -\gamma \dot{X},$ and the electrostatic force
is given by
$$
  F_\epsilon(X,V,Q)=\frac{1}{2}\frac{dQ_L}{dX}V_L + \frac{1}{2}\frac{dQ_R}{dX}V_R -
  \frac{1}{2}Q\frac{dV}{dX},
$$
where $Q$ and $V_{L,R}$ can be calculated from the equivalent
circuit of \mbox{fig.  \ref{fig:model_system}b}. Finally, the van
der Waals force between the grain and the leads is derived from a
Lennard-Jones interaction potential for the individual
atoms~\cite{a:israelachvili_85} in the leads and the grain
respectively. For a geometry with a metallic sphere close to a
flat substrate one finds,
\begin{eqnarray}
  F_{vdW} &=& \frac{H_{a}}{6}
  \left( \frac{r}{ \left( \frac{L}{2}-r-X \right)^2}
    - \frac{r}{ \left( \frac{L}{2}-r+X \right)^2} \right)\nonumber\\
    & -& \frac{H_{r}}{180}
  \left( \frac{r}{ \left( \frac{L}{2}-r-X \right)^8}
    - \frac{r}{ \left( \frac{L}{2}-r+X \right)^8 } \right) \nonumber,
\end{eqnarray}
where $H_{a,r}$ are the Hamaker constants for the attractive and
repulsive parts of the force. Equation (\ref{eq:Newton}) together
with eq.~(\ref{eq:rates}) describe the dynamics of the system.

\section{Features of the noise spectrum}
\begin{figure}
\begin{center}
\includegraphics[width=9cm]{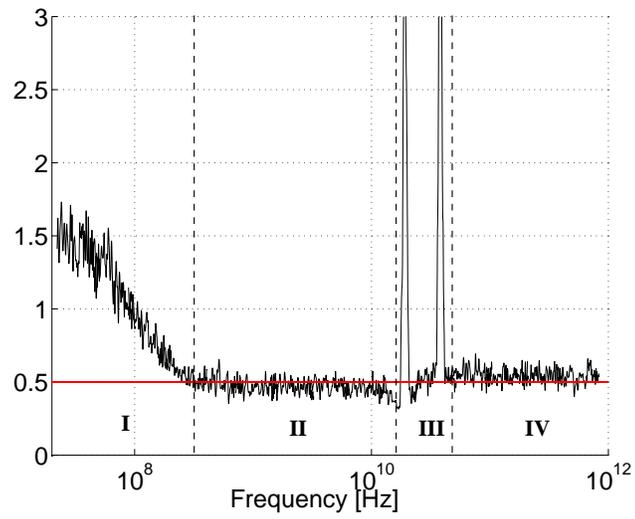}\caption{
  Power spectrum $S_{II}(\omega)$ for large amplitude shuttling.  For frequencies
  above the vibrational frequency the Fano factor is close to
  $1/2$ as for a static Coulomb blockade junction.  The peaks are
  located at the frequency of vibration and at the first harmonic. For
  frequencies below the vibrational frequency, the in time correlation
  due to the periodic grain motion leads to a slight suppression of
  the noise level. At still lower frequencies the the noise is
  increased due to slow fluctuations in oscillation amplitude.
  \label{fig:spectrum}}
  \end{center}
  \end{figure}

In this section we will briefly describe the current
noise-spectrum of the model system presented in the previous
section. Defining the current $I$ through the system as the charge
transferred per unit time from the grain to the right lead, the
PSD is given by
$$S_{II}(\omega)\equiv 2\int_{-\infty}^{+\infty} dt e^{i\omega t}\left<\Delta I(t)\Delta I(0)\right>$$
where the brackets denotes ensemble averaging.~\cite{Jong} By
doing direct numerical integration, {\it i.e.} solving the
stochastic differential equation eq.~(\ref{eq:Newton}), the
current as a function of time is obtained and the noise spectrum
calculated. A representative result of such a calculation is shown
in fig.~\ref{fig:spectrum}. This spectrum was calculated for a
system in the shuttle regime. The PSD has been normalized to
obtain the Fano factor $F={S_{II}}/({2e\left<I\right>}).$

We have divided the spectrum in fig.~\ref{fig:spectrum} into four
regions marked {\bf I}-{\bf IV}. At high frequencies, region {\bf
IV}, the Fano factor is close to a value of 1/2 which is the value
for a static double junction~\cite{Jong}. In region {\bf III} two
strong peaks are located at the vibration frequency and the first
harmonic. This is a result of the periodic charging and decharging
of the oscillating grain. The large magnitude of the first
harmonic stems from the fact that charge exchange between the
grain and the left lead gives rise to a displacement current in
the right lead where we measure $I$. Directly below the peaks,
region {\bf II}, the noise is suppressed below the shot noise
level of a static double junction, possibly due to the additional
in time correlations between successive tunnel events induced by
the oscillating grain.

The most interesting part of the spectrum however, is the low
frequency part in region~{\bf I}, where the Fano factor increases.
This enhancement of the noise is associated with fluctuations in
mechanical energy. In the next section we present an analytical
treatment of a simplified model system which does not take into
account non-linear elastic forces, vdW forces or position
dependent capacitances. Nevertheless, this model suffices to
explain the low frequency behavior of the noise spectrum in the
shuttle regime as well as to predict a (quasi) singular behavior
of the Fano factor at the instability threshold.

\section{Low frequency noise for weak electromechanical coupling}
The system is completely described by the conditional probability
densities $p(X,\dot{X},Q,t|X_0,\dot{X}_0,Q_0,t_0)$ to find the
system with charge $Q$ at time $t$ in the interval $[X,X+dX),
[\dot{X},\dot{X}+d\dot{X})$ given it was located around $(X_0,
\dot{X}_0,Q_0)$ at time $t_0$. The time evolution of the
conditional probability density, as well as the unconditional
probability density $p(X,\dot{X},Q,t)$, is given by the phase
space equation
\begin{eqnarray}
  \frac{\partial}{\partial t}p(X,\dot{X},Q,t)&=
  &-\frac{\partial}{\partial{X}}\left(\dot{X}p(X,\dot{X},Q,t)\right)\nonumber\\
  &+&\frac{\partial}{\partial\dot{X}}\left(\frac{\gamma\dot{X}-kX+F(Q)}{M}p(X,\dot{X},Q,t)\right)\nonumber
  \\
  &+&\sum_{Q^\prime\neq Q}\Gamma_{Q^\prime\rightarrow
    Q}(X)p(X,\dot{X},Q^\prime,t)\nonumber\\
  &-&\sum_{Q^\prime\neq Q}\Gamma_{Q\rightarrow
    Q^\prime}(X)p(X,\dot{X},Q,t).
    \label{eq:kolm}
\end{eqnarray}
Here a simplified version of the external force $F_{\rm
Ext.}=\gamma\dot{X}-kX+F(Q)$ where $F(Q)\equiv
F_{\epsilon}(X=0,Q,V)$ has been used. Introducing action-angle
variables $(E,\phi)$ defined through
$(X,\dot{X})=\sqrt{2E/M\omega^2}\left(\sin(\phi),\omega\cos(\phi)\right),$
where $\omega^2\equiv k/M$, eq.~(\ref{eq:kolm}) takes the form
  \begin{eqnarray}
    \frac{\partial}{\partial
      t}p(E,\phi,Q,t)&=&-\frac{\partial}{\partial{\phi}}\left(\left[\omega+\frac{\gamma}{M}
      \sin\phi\cos\phi-\frac{F(Q)\sin\phi}{\sqrt{2ME}}\right]p(E,\phi,Q,t)\right)
    \nonumber \\
    &-&\frac{\partial}{\partial{E}}\left(\left[-2\frac{\gamma}{M}
        E\cos^2\phi+\sqrt{\frac{2E}{M}}F(Q)\cos(\phi)\right]p(E,\phi,Q,t)\right)\nonumber
    \\&+&\sum_{Q^\prime\neq Q}\Gamma_{Q^\prime\rightarrow
      Q}(E,\phi)p(E,\phi,Q^\prime,t)\nonumber\\
      &-&\sum_{Q^\prime\neq
      Q}\Gamma_{Q\rightarrow
      Q^\prime}(E,\phi)p(E,\phi,Q,t).\label{eq:chapman}
  \end{eqnarray}

  To facilitate an analytical treatment it is convenient to rewrite eq.~(\ref{eq:chapman}). We thus introduce
  the dimensionless quantities
  $\tau=t\omega$, $n=Q/e$, $f_n=F(ne)/F(e)$,
  ${\cal
    E}=E/(\frac{1}{2}M\omega^2\lambda^2)$,
  $\tilde{\Gamma}_{n\rightarrow
    n^\prime}=\omega^{-1}{\Gamma}_{n\rightarrow n^\prime}$,
  $\epsilon=F(e)/k\lambda$, and $\eta=\gamma/(M\omega\epsilon)$.
The probability density can be written in vectorial form as ${\bf
P}({\cal E},\phi,\tau)$, where the $i-th$ component is given by
$p({\cal E},\phi,n(i),\tau)$ with $n(i)\equiv(-1)^{i-1}{\rm
Int}\left[\frac{i}{2}\right]$. We also introduce the matrices
$$
\hat{I}_{i,j}=\delta_{i,j},\,\,
\hat{f}_{i,j}\equiv\delta_{i,j}f_{n(i)},\,\,
\hat{\Gamma}^+_{i,j}=\tilde{\Gamma}_{n(j)\rightarrow n(i)},\,\,
\hat{\Gamma}^-_{i,j}=\delta_{i,j}\sum_{n^\prime}\tilde{\Gamma}_{n(i)\rightarrow
  n^\prime},
$$
and, define three operators
$\hat{\cal{L}}_1=-\frac{\partial}{\partial\phi}+\hat{\Gamma}^+({\cal
  E},\phi)-\hat{\Gamma}^-({\cal E},\phi)$,\\
$\hat{\cal{L}}_2=\frac{\partial}{\partial\phi}\left[\frac{\hat{f}}{\sqrt{{\cal
        E}}}\sin\phi-\hat{I}\eta\sin\phi\cos\phi\right]$, and
$ \hat{\cal L}_3=2\frac{\partial}{\partial {\cal
    E}}\left[\hat{I}\eta{\cal E}\cos^2\phi-\hat{f}\sqrt{{\cal
      E}}\cos\phi\right]$.
Equation~(\ref{eq:chapman}) can then be compactly written as
$$
\frac{\partial}{\partial \tau}{\bf P}({\cal
  E},\phi,\tau)=\left[\hat{\cal L}_1+\epsilon\left(\hat{\cal
      L}_2+\hat{\cal L}_3\right)\right]{\bf P}({\cal E},\phi,\tau).
$$
Since the low frequency noise is governed by a time scale much
longer than the period of vibration, adiabatic elimination of fast
variables can be used to treat slow fluctuations in amplitude.
This amounts to doing perturbation theory in $\epsilon$.

Following ref.~\cite{gardiner} we let ${\bf P}_\lambda$ and ${\bf
Q}_\lambda$ be normalized eigensolutions to $ \hat{\cal L}_1{\bf
P}_\lambda=\lambda({\cal E}) {\bf
  P}_\lambda$ and $\hat{\cal L}_1^\dagger{\bf
  Q}_\lambda=\lambda^*({\cal E}) {\bf Q}_\lambda,
$ corresponding to eigenvalue $\lambda$. ${\bf P}_0({\cal
E},\phi)$ is the stationary solution to the unperturbed problem,
i.e., $\hat{\cal L}_1{\bf P}_0=0$, while ${\bf Q}_0$ solves the
adjoint equation $\hat{\cal L}_1^\dagger {\bf Q}_0=0$. The
projector $\hat{\cal P}_0$ onto this state acting on an arbitrary
vector ${\bf x}({\cal E},\phi,\tau)$ is given by
$$
\left(\hat{\cal P}_0{\bf x}\right)({\cal E},\phi,\tau)={\bf
  P}_0({\cal E},\phi)\int_0^{2\pi}d\phi^\prime {\bf Q}_0^*({\cal
  E},\phi^\prime)\cdot{\bf x}({\cal E},\phi^\prime,\tau).
$$
By direct insertion, using that $\hat{\cal
L}^\dagger_1=\frac{\partial}{\partial\phi}+\hat{\cal G}^\top({\cal
  E},\phi)$, one finds that $({\bf Q}_0)_n=[1, 1,\cdots,1]^\top$.
Next, splitting up ${\bf P}({\cal E},\phi,\tau)$ in mutually
orthogonal parts $ {\bf v}\equiv \hat{\cal P}_0{\bf P}$ and ${\bf
w}\equiv (1-\hat{\cal P}_0){\bf P}$, an equation, correct to
second order in $\epsilon$, can be found for ${\bf {v}}$;
\begin{equation}
\frac{\partial}{\partial \tau}{\bf v}=\epsilon\hat{\cal
  P}_0\hat{\cal L}_4{\bf v}-\epsilon^2\hat{\cal P}_0\hat{\cal
  L}_3\hat{\cal L}_1^{-1}[\hat{\cal L}_2+\hat{\cal L}_3-\hat{\cal
  P}_0\hat{\cal L}_4]{\bf v}.
\label{eq:veq}
\end{equation}
Here $\epsilon\hat{\cal{L}}_4=\hat{I}\frac{\partial }{\partial
{\cal
    E}}\left[\gamma({\cal E})-{\cal W}({\cal E})\right]$, which
 contains the average dissipation per cycle $\gamma({\cal E})$
and the average pumped energy per cycle ${\cal
  W}({\cal E})$ defined through
$$
\gamma({\cal E})\equiv {2\epsilon\eta}{\cal
  E}\int_0^{2\pi}d\phi{\bf Q}_0^*\cdot {\bf P}_0({\cal
  E},\phi)\cos^2\phi,\,\,
{\cal W}({\cal E})=2\epsilon\sqrt{\cal E}\int_0^{2\pi}d\phi {\bf
Q}_0^*\cdot \hat{f}{\bf P}_0({\cal E},\phi)\cos\phi.
$$
Letting $\rho({\cal E},\tau)$ be the probability density for
finding the system with mechanical energy $\cal E$, {\it i.e.}
${\bf v}={\bf P}_0{\rho}$, and using eq.~(\ref{eq:veq}) a
Fokker-Planck equation for $\rho$ is obtained;
\begin{eqnarray}
  \frac{\partial}{\partial \tau}\rho({\cal
    E},\tau)&=&-\frac{\partial}{\partial {\cal E}}[({\cal W}({\cal
    E})-\gamma({\cal E})+{\cal O}(\epsilon^2))\rho({\cal
    E},\tau)]\nonumber\\
  &-&\epsilon^2\frac{\partial^2}{\partial {\cal
      E}^2}\left(\sum\limits_{\lambda\neq 0}\frac{f_\lambda({\cal
      E})g_\lambda({\cal E})}{\lambda({\cal E})}\rho({\cal
      E},\tau)\right),\nonumber
\end{eqnarray}
where $f_\lambda({\cal E})\equiv \int_0^{2\pi}d\phi {\bf
Q}_0^*\cdot\hat{\cal O}{{\bf P}_\lambda({\cal E},\phi)}$ and
$g_\lambda({\cal E})\equiv \int_0^{2\pi}d\phi {\bf
Q}_\lambda^*\cdot\hat{\cal O}{\bf P}_0({\cal E},\phi),$ with\\
$\hat{\cal O}\equiv\left[{2\eta{\cal
E}}\cos^2\phi-2\hat{f}\sqrt{\cal E}\cos\phi\right]$. Since $({\cal
W}({\cal E})-\gamma({\cal E}))$ is of order $\epsilon$ the small
energy shift of order $\epsilon^2$ can be ignored to a first
approximation and an equivalent Ito stochastic differential
equation for the vibrational energy ${\cal E}$ is arrived at
$$d{\cal E}=\left[{\cal W}({\cal E})-\gamma({\cal
      E})\right]d\tau+\alpha({\cal E})dW(\tau) \label{eq:ito},$$
where ${W}(\tau)$ is the Wiener process and $\alpha({\cal
E})\equiv \epsilon\sqrt{-2\sum_{\lambda\neq
    0}f_\lambda({\cal E})g_\lambda({\cal E})}.
$

Performing a small noise expansion~\cite{gardiner} around ${\cal
E}_0$, where ${\cal E}_0$ solves $({\cal W}({\cal
  E}_0)-\gamma({\cal E}_0))=0$, an
analytical expression for the PSD can be obtained
\begin{equation}
  S_{II}(\omega)=2\frac{[\alpha({\cal E}_0){I}^\prime({\cal
      E}_0)]^2}{(\gamma^\prime({\cal E}_0)-{\cal W}^\prime({\cal
      E}_0))^2+\omega^2} \label{eq:lfnoise}.
\end{equation}
Here the current $I$ is given by $ I({\cal
E})=\frac{e\omega}{2\pi}\int_0^{2\pi}d\phi {\bf
  Q}_0^*\cdot\hat{\cal J}({\cal E},\phi) {\bf P}_0({\cal E},\phi).
$ where $ \left(\hat{\cal J}({\cal E},\phi)\right)_{i,j}\equiv
\delta_{i,j}\tilde{\Gamma}_{n(i)\rightarrow n(i)-1}^{\rm R}({\cal
  E},\phi) -\tilde{\Gamma}_{n(i)\rightarrow n(i)+1}^{\rm R}({\cal
  E},\phi)
$ is the current operator.

Equation (\ref{eq:lfnoise}) is our central result. Two important
conclusions can be drawn from this equation. First, below a break
frequency given by $\omega_c=(\gamma^\prime({\cal E}_0)-{\cal
W}^\prime({\cal E}_0))$ the noise will rise due to fluctuations in
oscillation amplitude. This agrees with the PSD in
fig.~\ref{fig:spectrum}. Second, approaching the transition point
(from above) $(\gamma^\prime({\cal E}_0)-{\cal W}^\prime({\cal
E}_0))\rightarrow 0$ (see ref.~\cite{a:98_isacsson}) which implies
that the low frequency noise may display a divergent behavior at
the point of transition from shuttle to stationary regime.

\begin{figure}
  \begin{center}
  \includegraphics[width=8cm]{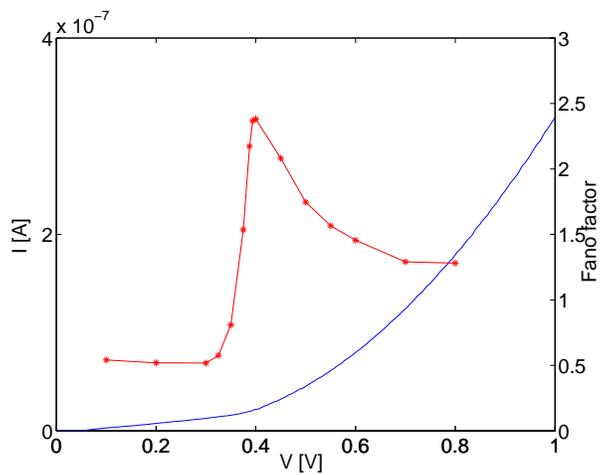}
   \caption{Current
voltage characteristics plotted together with
  $S_{II}(\omega\rightarrow 0)$. The current is the solid line with the scale on the left ordinate
  while the Fano factor is shown for a discrete set of points
  with the scale on the right ordinate (Lines connecting the points are a guide to the eye). Below the critical voltage where
  there is no sustained grain motion the Fano factor is that of a Coulomb blockade double junction. Above the critical voltage
  the grain is oscillating and the Fano factor is increased and
  shows a divergent behavior at the critical voltage in accordance with
  eq.~(\ref{eq:lfnoise}).\label{fig:iv}}
\end{center}
\end{figure}

To investigate this eqs.~(\ref{eq:rates}-\ref{eq:Newton}) were
simulated for a sequence of voltages for one typical system and
the resulting $I-V$ curve as well as the Fano factor are shown in
fig.~\ref{fig:iv}. The parameter values used are shown in
table~\ref{tab:table1}. They correspond to a nanometer sized Au
grain commonly used in experiments with self-assembled Coulomb
blockade double junctions.

Although, as explained in~\cite{a:04_nord_isacsson}, the
non-parabolic confining potential smoothens any step-structure in
the current-voltage characteristics, the transition between
static- and shuttle-operation is clearly visible in the noise
spectrum. Below the threshold voltage the  Fano factor is of the
order 1/2 whereas in the shuttle regime, it is higher. In
accordance with eq.~\ref{eq:lfnoise}, approaching the threshold
from above (higher to lower voltages) the PSD shows divergent
behavior.

\section{Conclusions}
Using adiabatic elimination of fast variables and numerical
integration of eqs.~(\ref{eq:rates}-\ref{eq:Newton}) the PSD of
the classical single electron shuttle has been studied in the case
of weak electromechanical coupling. We have focussed on the low
frequency part of the spectrum, which is the part most susceptible
to direct measurements, and found that the shuttle regime can be
distinguished from the stationary regime by the noise level at low
frequencies.

In particular, the shuttle regime is associated with a rise in the
low frequency PSD, as compared to a stationary Coulomb blockade
double junction, resulting in an increased Fano factor. This
increase is due to slow variations in the current arising from
variations in oscillation amplitude. Approaching the point of the
transition from above (lowering the bias voltage when in the
shuttle regime) the noise level shows a (quasi) divergent behavior
upon closing in on the transition. Hence, even though a
measurement of the average current alone may not reveal whether
the system is in the shuttle regime or not, the accompanying noise
signature can provide this information.

\section{acknowledgments}
  The authors are grateful for stimulating discussions with R.
  I.  Shekhter and L. Y. Gorelik. This work has received financial
  support (A.I.) through
  The Swedish Foundation for International Cooperation in Research and
  Higher Education (STINT), and (T.N.) the Swedish Research Council (VR).

\newpage
\newpage
\clearpage

\clearpage
\newpage
\newpage
\clearpage
\begin{center}
\begin{table}
\caption{Numerical values of parameters used to obtain data of
fig.~\ref{fig:iv}.\label{tab:table1} } \vspace*{1cm}
\begin{tabular}{lcr|lcr|lcr}
Quantity&Value&Units&Quantity&Value&Units&Quantity&Value&Units\\
\hline $L$                   &  $5$  & $nm$     & $k$     &  $1$ &
$N/m$ & $\lambda$             &  $0.1$& $nm$  \\
$r$                   &  $1$  & $nm$     & $X_L$,$X_R$    &  $1$ &
$nm$ &
$R_0^L$,$R_0^R$       &  $10$ & $M\Omega$ \\
$M$     &  $10^{-23}$ & $kg$ &  $H_{a}$  & $4\cdot10^{-19}$ & $Nm$
& $C_0^L$,$C_0^R$,$C_0$ &  $1$  & $aF$\\
$\gamma$ &  $10^{-13}$ & $kg/s$ & $H_{r}$ & $10^{-72}$ &$Nm^7$ &
$A_L$,$A_R$      & $2.5$ & $nm$ \\
\end{tabular}
\end{table}
\end{center}
\end{document}